\documentclass[showpacs,preprintnumbers,amsmath,amssymb,12pt,floatfix,epsfig,nofootinbib]{revtex4}
\usepackage{amssymb}
\usepackage{bm}
\usepackage{graphicx}
\newcommand{\be}{\begin{eqnarray}}
\newcommand{\ee}{\end{eqnarray}}

\begin{document}
\draft
\title{Neutron stars in a perturbative $f(R)$ gravity model with strong magnetic fields}

\author{Myung-Ki Cheoun}
\thanks{cheoun@ssu.ac.kr}
\affiliation{Department of Physics, Soongsil University, Seoul 156-743, Korea}

\author{Cemsinan Deliduman}
\thanks{cemsinan@msgsu.edu.tr}
\affiliation{Department of Physics, Mimar Sinan Fine Arts University,
Bomonti 34380, Istanbul, Turkey}

\author{Can G\"{u}ng\"{o}r, Vildan Kele\c{s}}
\affiliation{\.{I}stanbul Technical University, Faculty of
Science and Letters,\\ Physics Engineering Department,
Maslak 34469, Istanbul, Turkey}

\author{C. Y. Ryu}
\affiliation{General Education Curriculum Center, Hanyang University,
Seoul 133-791, Korea}

\author{ Toshitaka Kajino}
\affiliation{National Astronomical Observatory of Japan, 2-21-1 Osawa,
Mitaka, Tokyo 181-8588, Japan \\
Department of Astronomy, Graduate School of Science, University of
Tokyo, Hongo 7-3-1, Bunkyo-ku, Tokyo 113-0033, Japan }

\author{Grant J. Mathews}
\affiliation{Center for Astrophysics, Department of Physics, University of Notre Dame,
IN 46556, USA}

\begin{abstract}
We investigate the effect of a strong magnetic field on the structure of neutron stars in a model with perturbative $f(R)$ gravity. The effect of an interior strong magnetic field of about $10^{17 \sim 18}$ G on the equation of state is derived in  the context of a quantum hadrodynamics (QHD) model. We solve the modified spherically  symmetric  hydrostatic equilibrium equations derived for a gravity model with  $f(R)=R+\alpha R^2$. Effects of both the finite  magnetic field and the modified gravity are detailed for various values of the magnetic field and the perturbation parameter $\alpha$ along with a discussion of their physical implications. We show that there exists a parameter space of the modified gravity and the magnetic field strength, in which even a soft equation of state can accommodate a large ($> 2$ M$_\odot$) maximum neutron star mass through the modified mass-radius relation.
\end{abstract}


\maketitle

\section{Introduction}

Neutron stars, as a remnant of  supernova explosions, are an excellent probe of  nuclear matter in extreme environments.   Among the observed neutron-star phenomena, soft $\gamma$-ray repeaters (SGRs) and anomalous $X$-ray pulsars (AXPs) are believed to provide evidence for magnetars \cite{Duncan92,Thompson96}, {\it i.e.} neutron
stars with strong surface
magnetic fields of $10^{14} \sim 10^{15}$ G (see \cite{Mereghetti08} for a review).
In the interior of these magnetic neutron stars, the magnetic
field strength could be as high as $10^{18}$ G according to the scalar
virial theorem. Such strong magnetic fields may affect properties
of neutron stars, for example, the relative populations of various particles, the equation of
state, and the mass-radius relation. Many studies of dense nuclear
 matter in the presence of strong magnetic fields have been reported \cite{cardall2001,band1997,brod2000,su2001,Brod02,dey2002,shen2006,shen2009,panda2009,Rabbi10,Ryu10}. These
works have considered the electromagnetic interaction,
the Landau quantization of charged particles, and the baryon anomalous
magnetic moments (AMMs).
Indeed the detailed analysis of neutron stars with strong magnetic
fields is an active  area of current research.

Of relevance to the present work are separate studies in which  various  modifications of Einstein's general relativity have been introduced to explain the current accelerating cosmic expansion. One can simply introduce a cosmological constant to explain this accelerated expansion, but the theoretical prediction for the  value of the cosmological constant from  quantum field theory is many orders of magnitude larger than the value  inferred from the astronomical data. Equivalently one can postulate  negative-pressure  energy-momentum contribution of unknown origin, i.e. the so called dark energy component with $\rho/p \simeq - 1$, to the matter side of the Einstein equation. An alternative to these  approaches, however,  is to modify the geometry side of the Einstein equation.  Such modifications can arise from so called modified gravity theories. A class of modified gravity theories is the $f(R)$ gravity (see reviews
\cite{Odintsov-rev,Capozziello-rev,Sotiriou-rev,deFelice-rev} and references therein).  This modified gravity is unique for its simplicity and has been shown to be compatible with the constraints from terrestrial  laboratory measurements, along with the Solar System constraints and neutron star tests \cite{Joa10,Cemsi11,Cooney,Orellana}.

\section{Modified Gravity}

The $f(R)$ gravity theory is simply defined by the following minimal modification to the Einstein-Hilbert action,
\begin{equation}
S = { 1 \over { 16 \pi}} \int d^4 x \sqrt{ -g} f (R) + S_{\rm matter},
\end{equation}
where $g$ denotes the determinant of the metric $g_{\mu \nu}$ and $R$ is the Ricci scalar. Here we set $G = 1$ and $c = 1$. In the present work we assume that the function $f(R)$ has the following linearized perturbative form with respect to a small parameter $\alpha$ without a cosmological constant term,
\begin{equation}
f(R) = R + \alpha h(R)~,
\label{fReq}
\end{equation}
where $h(R)$ is an arbitrary function of $R$. The modified TOV equations in this gravity are then given by \cite{Cemsi11}:
\begin{eqnarray}
  \frac{dM_\alpha}{dr} &=& 4\pi r^{2} \rho_\alpha -\alpha h_R \left[
  \begin{array}{l}
    4\pi r^{2}\rho  +\frac{r^2}{4} (\frac{h}{h_R}-R) \\
    +(2\pi\rho r^3-r+\frac32 M)\frac{h_R'}{h_R}-\frac12 r(r-2M)\frac{h_R''}{h_R}
  \end{array}
    \right] \label{1stTOV}~,
\end{eqnarray}
\begin{eqnarray}
&& {{d P_{\alpha}  } \over { dr  }} = - ( \rho_{\alpha} + P_{\alpha}) {  {d \phi_{\alpha}  } \over { dr }}~,
 \nonumber \\
&&  ( r - 2M_{\alpha} ) {  {d \phi_{\alpha}  } \over { dr }} = 4 \pi r^2 P_{\alpha}
+ { M_{\alpha} \over r} - \alpha h_R \left(
\begin{array}{l}
4 \pi r^2 P + {r^2 \over 4} ( { h \over h_R} - R) \\
+ ( r - \frac32 M + 2 \pi P r^3 ) {  {h^{'}_R  } \over { h_R }}
\end{array}
\right) ~,\label{2ndTOV}
\end{eqnarray}
where $\rho_{\alpha} = \rho + \alpha \rho_1 + ...$, and similarly
$P_{\alpha}$ and $M_{\alpha}$ are expanded in terms of $\alpha$, {\it i.e.} $P_{\alpha} = P + \alpha P_1 + ...$ and $M_{\alpha} = M + \alpha M_1 + ...$ with its zeroth order $M= 4 \pi \int \rho (r) r^2 dr$. $h_R (R)$ is the first derivative of $h(R)$ with respect to $R$, whereas $h'_R (R)$ and $h''_R (R)$ are the first and the second derivatives of
$h_R (R)$ with respect to $r$.
If we take the $\alpha = 0$ limit, this equation reduces to the standard TOV equation of general relativity.
It should be noted that we set
$M_\alpha (M) \rightarrow 2M_\alpha (2M)$ compared to the modified TOV equations in \cite{Cemsi11}.

The value of $\alpha$ may be constrained by the observational data. For example, with a similar analysis as is done in the present work it has been found in \cite{Cemsi11} that $\alpha \le 2 \times 10^{5}\; {\rm m}^2$.  For Gravity Probe B it has been deduced that  $\alpha \le 5 \times 10^{11}\; {\rm m}^2$ \cite{Joa10}, while  for the terrestrial E{\"o}t-Wash experiment the inferred constraint is  $\alpha \le 10^{-10}\; {\rm m}^2$ \cite{Joa10}. All of this means that allowed values for the parameter $\alpha$ depend  heavily on the length scale considered. In fact, it can be argued \cite{Cemsi11} that a length scale given by $\sqrt{\alpha}$ has nearly the same order of magnitude as the typical dimension of the probes used in above tests.
Moreover, it is also related to the Yukawa correction to the Newtonian potential, ${G \over 3}~ \exp( - r / \lambda)$ with a length scale parameter $\lambda = \sqrt{ 6 \alpha}$ \cite{Joa10}. Such a violation of the inverse-law of Newtonian gravity has been suggested as a possible means to resolve a current dilemma in  neutron star physics, namely the inconsistency of the super-soft equations of state \cite{BAO}, obtained by interpreting heavy ion collision data FOPI/GSI \cite{Fopi}, with the presently observed \cite{demorest2010} upper mass limit on neutron stars of $M \ge 1.97 \pm 0.04 $ M$_\odot$.

In the present study we consider the combination of strong magnetic fields and modified gravity.  By way of motivation for this we note that in the  five dimensional unification of gravity and electromagnetism the Kaluza--Klein action expands into:
\begin{equation}
\mathcal{R} \rightarrow f(R) = R - \alpha \vert F \vert^2 ~~,
\end{equation}
where $\mathcal{R}~(R)$ is the scalar curvature in five (four) dimensions, $F$ is the four dimensional electromagnetic field strength and in this case $\alpha$ relates to (the square of) the length scale of the extra dimension.  In this context, therefore, it is perhaps natural to compare a modification of general relativity together with a strong  electromagnetic field.  Since we do not consider charged neutron stars, the Maxwell stress tensor simply reduces to the energy density in the magnetic field.  Hence, in what follows we consider the effects of magnetic field and modified gravity on the neutron star structure combined and contrasted as a first step toward a unified picture.  As a simplification we consider modified gravity with $h(R) \sim R^2$ in Eq. (\ref{fReq}).  Technically, such a term is motivated by Lovelock or  Gauss--Bonnet gravity in higher dimensions, however, we keep in mind that this could also be loosely associated with the Kaluza--Klein electromagnetism.

\section{Equations of State and Numerical Analysis}

Many theoretical models for the equation of state (EoS) of the neutron star have been  developed starting from  from an  {\it ab initio} or effective nucleon-nucleon interaction in order to explain the observational  mass-radius relation data \cite{Lattimer12}. In this work, we exploit our previously developed  EoS \cite{Ryu10} based upon  a QHD Lagrangian and applied to describe the interior structure of magnetars. Detailed expressions have been given in that work and will not be repeated here.

Employing this EoS we  have numerically  integrated Eqs. (\ref{1stTOV}) and (\ref{2ndTOV}).
We start from the center of the star for a certain value of central pressure,
$P_{\mathrm{c}}$, and then utilize a Runge-Kutta scheme with a fixed step size of
$\Delta r=0.001$ km. The radius of the star, $R_\star$, is identified as
the point where pressure drops to a very small value ($\le 10$ dyne/cm$^{2}$).  At that point
we record the mass of the star, $M_\star$. We vary the central density,
$\rho_c$, from $2\times 10^{14}$ g cm$^{-3}$ to $1\times
10^{16}$ g cm$^{-3}$ (to $2\times
10^{16}$ g cm$^{-3}$ in some cases) in 200 logarithmically equal steps to obtain a sequence of equilibrium configurations. We record the mass and the radius for each central pressure. This allows us to construct the  mass-radius (M--R) relation for a given EoS. We have then repeated  this procedure for a range of values for $\alpha$ to quantify  the effect of the perturbative term added to the Lagrangian.


In the following, we present results for the effect of strong magnetic fields on neutron stars with a TOV solution based upon perturbative $f(R)$  gravity. The mass-radius relation and the mass vs.~central density for each EoS are given for 5 representative values of $\alpha_9 \equiv  \alpha / 10^9 \, {\rm cm^2} = -2,-1,0,1,2 $ in the $f(R)  = R + \alpha R^2$ gravity. The magnetic field strength inside the neutron star is assumed to obey a functional dependence on density given by \cite{band1997,shen2009}
\begin{equation}B( \rho / \rho_0 ) = B^{\rm surf} + B_0 [ 1 - \exp [( - \beta ( \rho / \rho_0 )^{\gamma}]]~.
\label{Mfield}
\end{equation}
Here, $B^{\rm surf}$ is the magnetic field at the surface taken as $10^{15}$ G from observations and $B_0$ is the saturation value of the interior  magnetic field at high densities. In the present work, we adopt a somewhat rapidly declining magnetic field strength with density ($ \beta = 0.02$ and $\gamma = 3$). Since the magnetic field is usually specified  in
units of the critical field for the electron, $B^c_e = m_e^2/e
= 4.414 \times 10^{13}\, {\rm G}$, the $B$ and the $B_0$ in Eq. (\ref{Mfield}) can be written as $B^* = B/B^c_e$ and $B_0^* = B_0/B^c_e$. In this work, we regard the $B_0^*$ as a free parameter.

Figure 1 shows the  mass-radius relations (left panels)  and the mass vs.~central density (right panels)  for the case of  a $np$ phase,  i.e.  no hyperons.  The magnetic field strength is taken to be $B_0^* \sim 10^{2 \sim 3}$ for  the 5 representative values of $\alpha_9$.
In order to constrain the value of $\alpha$, the recent measurements of the mass and radius of neutron stars, EXO 1745-248 \cite{oze1745}, 4U 1608-52 \cite{guv1608} and 4U 1820-30 \cite{guv1820} are used. The region bounded by the thin black line in all M--R plots is the $2\sigma$ confidence contour
based upon these three constraints \cite{Ozel10}. We also use the mass of PSR J1614-2230 with $1.97\pm 0.04\,M_{\odot }$ \cite{demorest2010} as a constraint.  This is shown as the horizontal black line with a grey error-bar. For a viable  EoS together with an allowed value of the perturbation parameter $\alpha$, the maximum neutron-star mass should lie  above this constraining contour. Within the framework of the modified gravity considered here these two constraints eliminate many of the possible equations of state.

The two upper-most figures show results without a magnetic field for the $np$ phase.  These figures illustrate  the effect of the modified gravity alone. Negative (positive) $\alpha_9$ values give rise to a stiff  (soft) EoS.
This highlights the  interesting feature  that modified gravity can  lead to a mass-radius relation that mimics a stiff or soft EoS for neutrons stars.  Indeed, if the EoS is ever established to be soft, modified gravity of the sort studied here may be required to explain neutron star masses as large as 2 $M_\odot$.

In particular, one may note that negative $\alpha$ values lead to larger masses beyond 2.0 $M_{\odot}$, while a positive $\alpha$ value tends to diminish the maximum  mass.  The behavior of the EoS in the high density region also has another interesting property in that it shows a very strong softness around the region with densities of $\sim 2 -3  \rho_c$. Magnetic field strengths of about $B_0^* \sim 10^{2 \sim 3}$ have little  effect  on the maximum mass  and mass-radius relations as shown in Figure 1. However,  with a larger magnetic field of $B_0^* \sim 10^{4}$, a stiff EoS  and a larger maximum mass  is obtained as shown in Figure 2. However, if we assume that the $\alpha_9$ values are positive, i.e.  considering the interpretation of a Yukawa correction to the Newtonian gravity,  the strong magnetic field effects for  $B_0^* \sim 10^{4}$ are compensated by the modified gravity effects.

The remaining figure 3 is  for the case of a  $nph$ phase that  includes hyperons. We observe exactly the same phenomena in this case, i.e. the  effects of modified gravity are seen as mimicking a stiff or soft EoS for  neutron stars depending upon whether $\alpha$ is negative or positive, respectively. In particular we observe that  smaller (and more negative) values of $\alpha$ allow for a  higher maximum   neutron star mass.
Therefore,  the perturbative parameter $\alpha$ of  modified gravity is a new degree of freedom that can  alter the M--R relation and (for negative $\alpha$) causes some equations of state to be viable for neutron star matter which might not have otherwise been allowed.

\section{Conclusions}

In summary, in this paper we have considered the effects of strong magnetic fields on the neutron star mass-radius relation by using a modified TOV equation derived from the $f(R) = R + \alpha R^2$ modification of Einstein's general relativity. We fixed the upper bound of the expansion parameter $\alpha_9 = \alpha/10^9 ~ {\rm cm}^2$ of the gravity model using several constraints based upon neutron star observations. It turns out that in the case of $\alpha_9 > 0$, effects by the modified gravity can be compensated by those of a strong magnetic field. However, the case of $\alpha_9 < 0$ is completely different: in this regime some equations of state  which were not viable for neutron stars  in the case of general relativity  become viable again. This is in accord with the previous results given in \cite{Cemsi11,Cemsi12} on the consequences of the perturbative modification of general relativity.

It would be instructive to repeat the present analysis for the case of a non-spherically symmetric, but axially symmetric space-time, because a high magnetic field could alter the almost spherically symmetric shape of a neutron star to an axially symmetric one. Then one would better understand the interplay between the high magnetic fields and the effects of gravity on the physics of neutron stars. In the case of white dwarfs with high magnetic fields, the magnetic field's possible effect on the star's geometry and the consequence for its mass and radius has been commented upon in \cite{Das13}. The present work should be considered as a first step toward that line of research.

\acknowledgements

C.D. thanks K. Y. Ek\c{s}i for valuable discussions and pointing out the last reference.
C.D. also thanks The National Astronomical Observatory of Japan for hospitality where collaboration for this work started.  This work was supported by the National Research Foundation of Korea (2011-0003188 and 2011-0015467), and also
supported in part by Grants-in-Aid for Scientific Research of JSPS (20244035). This work was also supported by the Grants-in-Aid for the Scientific Research
from the Ministry of Education, Science and Culture of
Japan (20244035, 21540412), Scientific Research on Innovative Area of MEXT (20105004), and Heiwa Nakajima Foundation. Work at the University of Notre Dame
(G.J.M.) supported by the U.S. Department of Energy
under Nuclear Theory Grant DE-FG02-95-ER40934.


\begin{figure}[H]
\includegraphics[width=5.5cm]{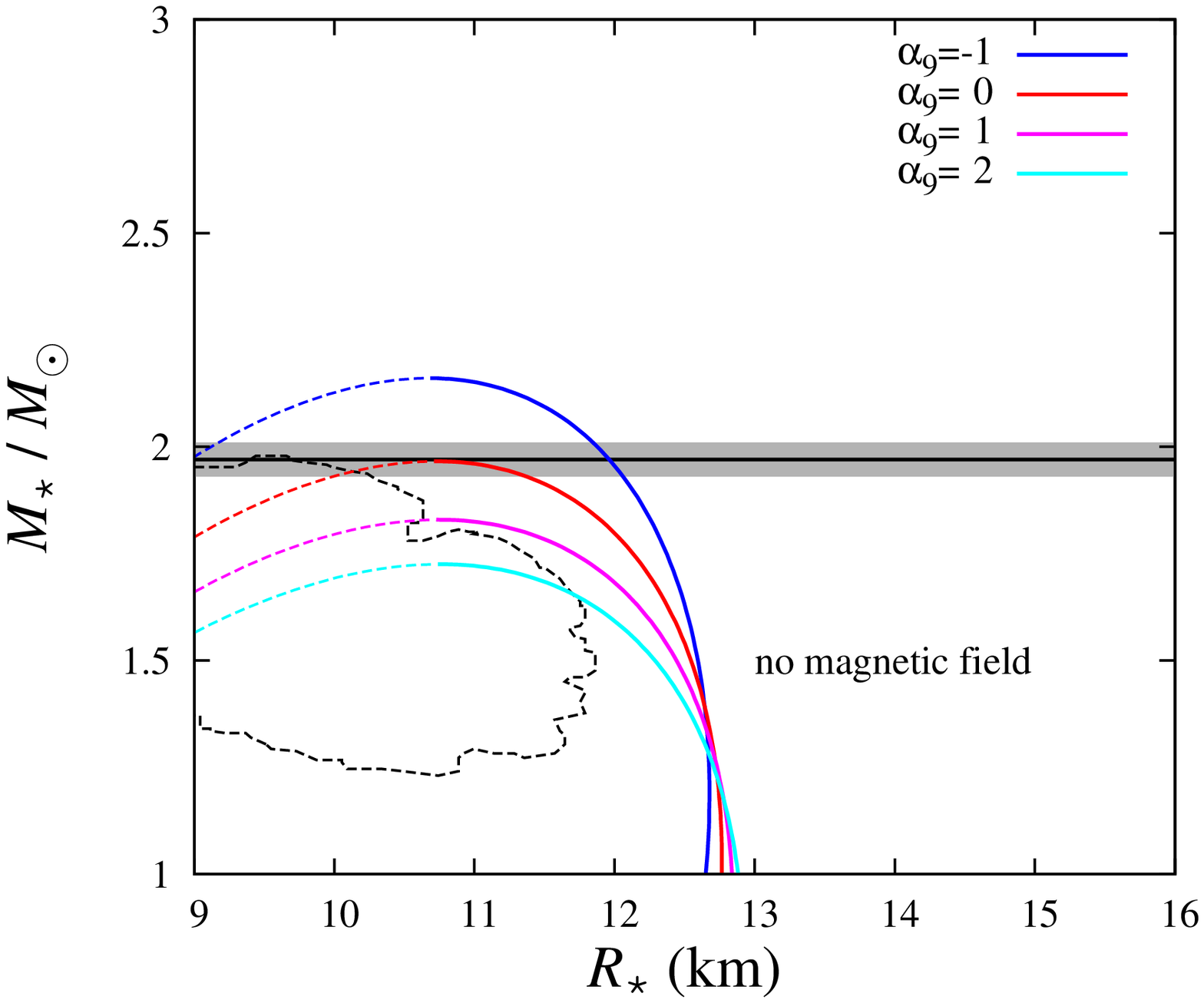}
\includegraphics[width=5.5cm]{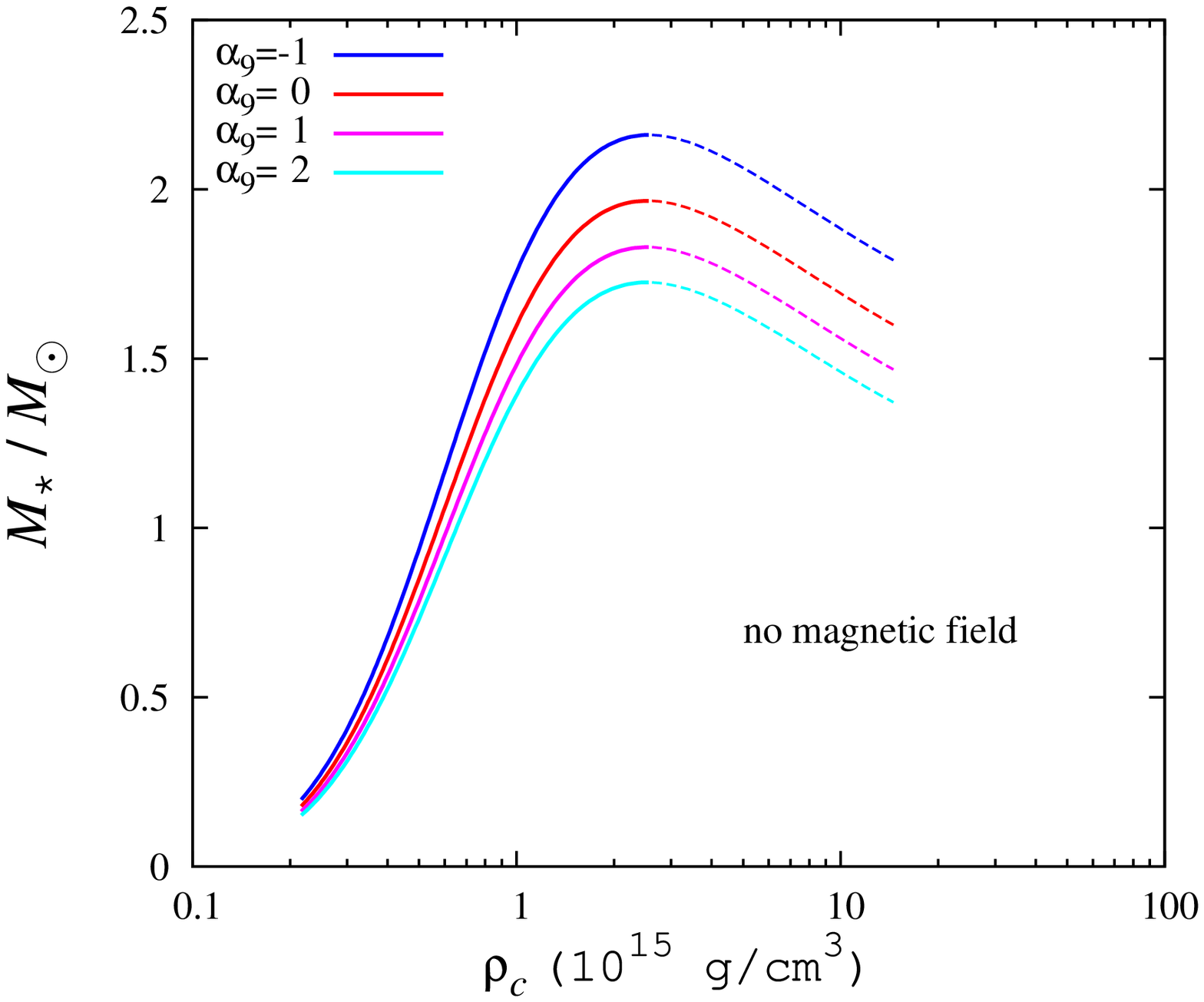}\\
\includegraphics[width=5.5cm]{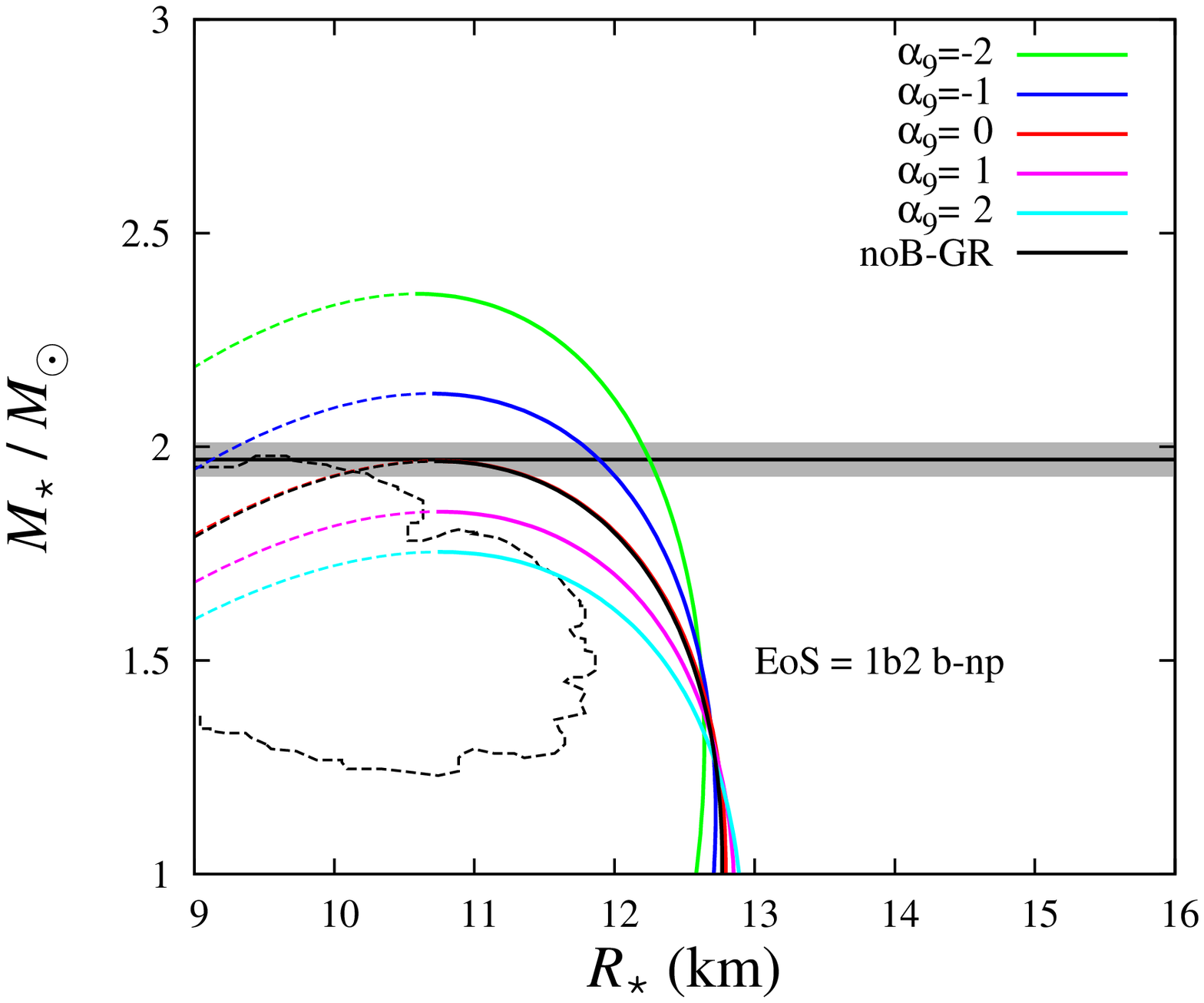}
\includegraphics[width=5.5cm]{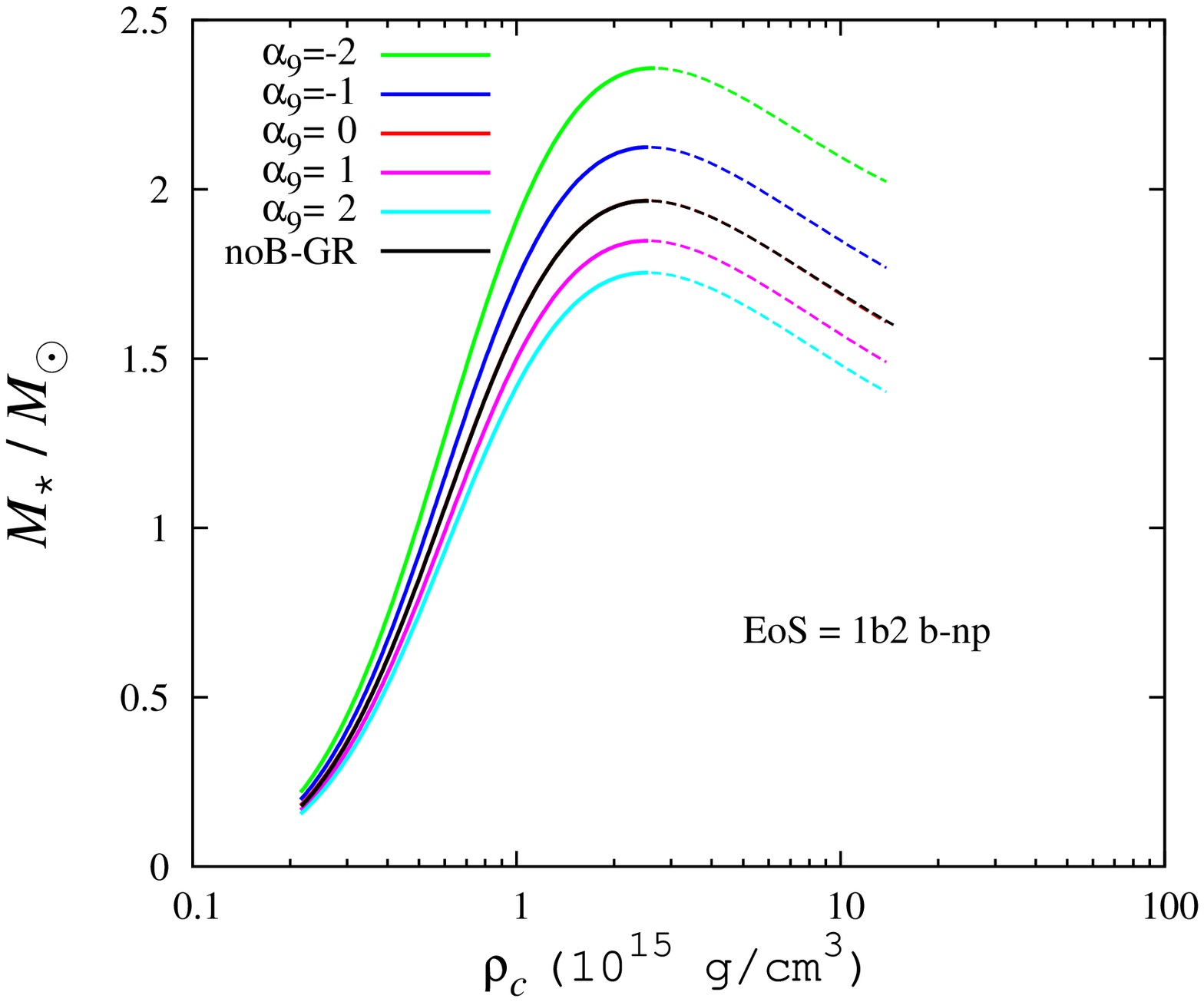}\\
\includegraphics[width=5.5cm]{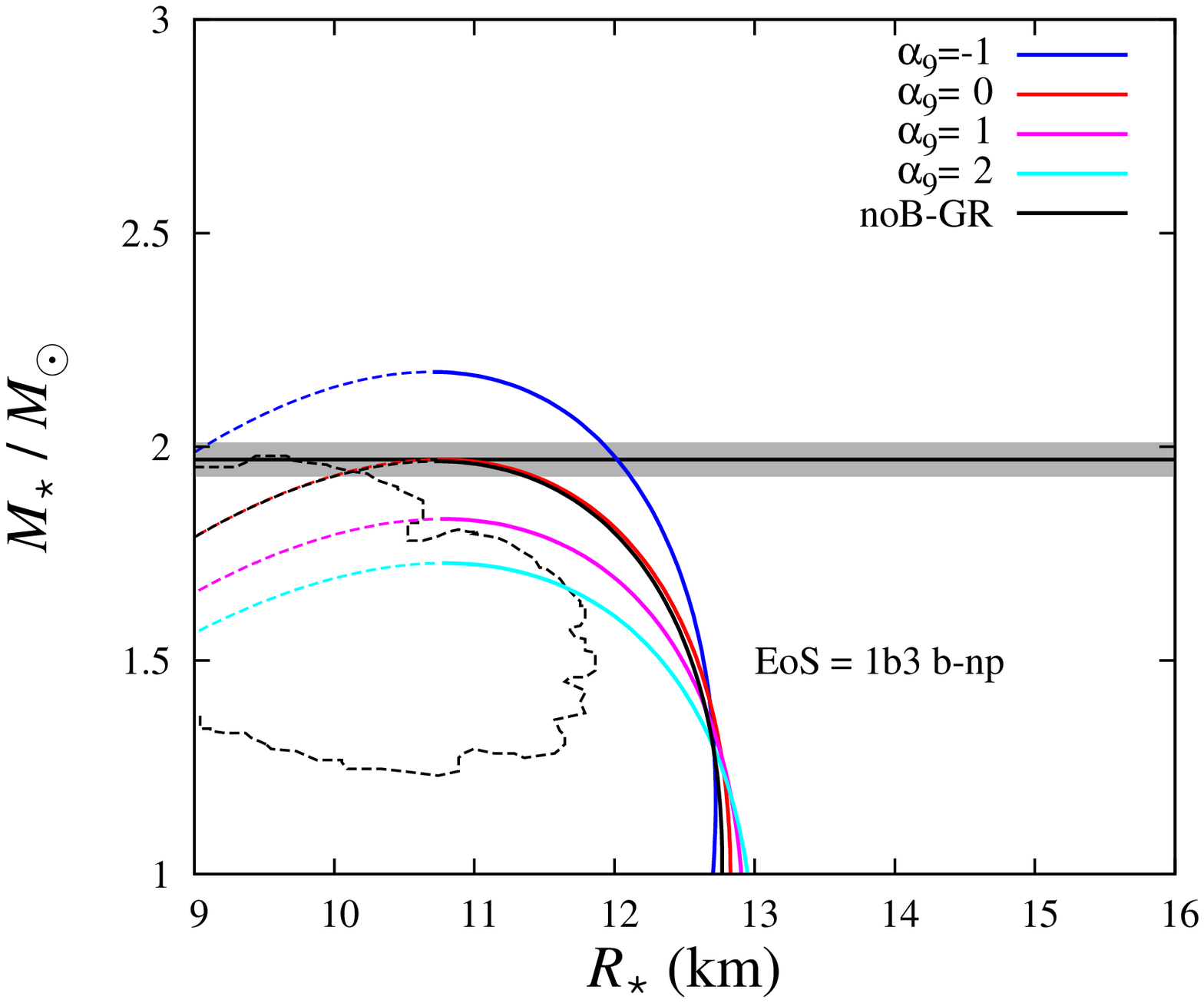}
\includegraphics[width=5.5cm]{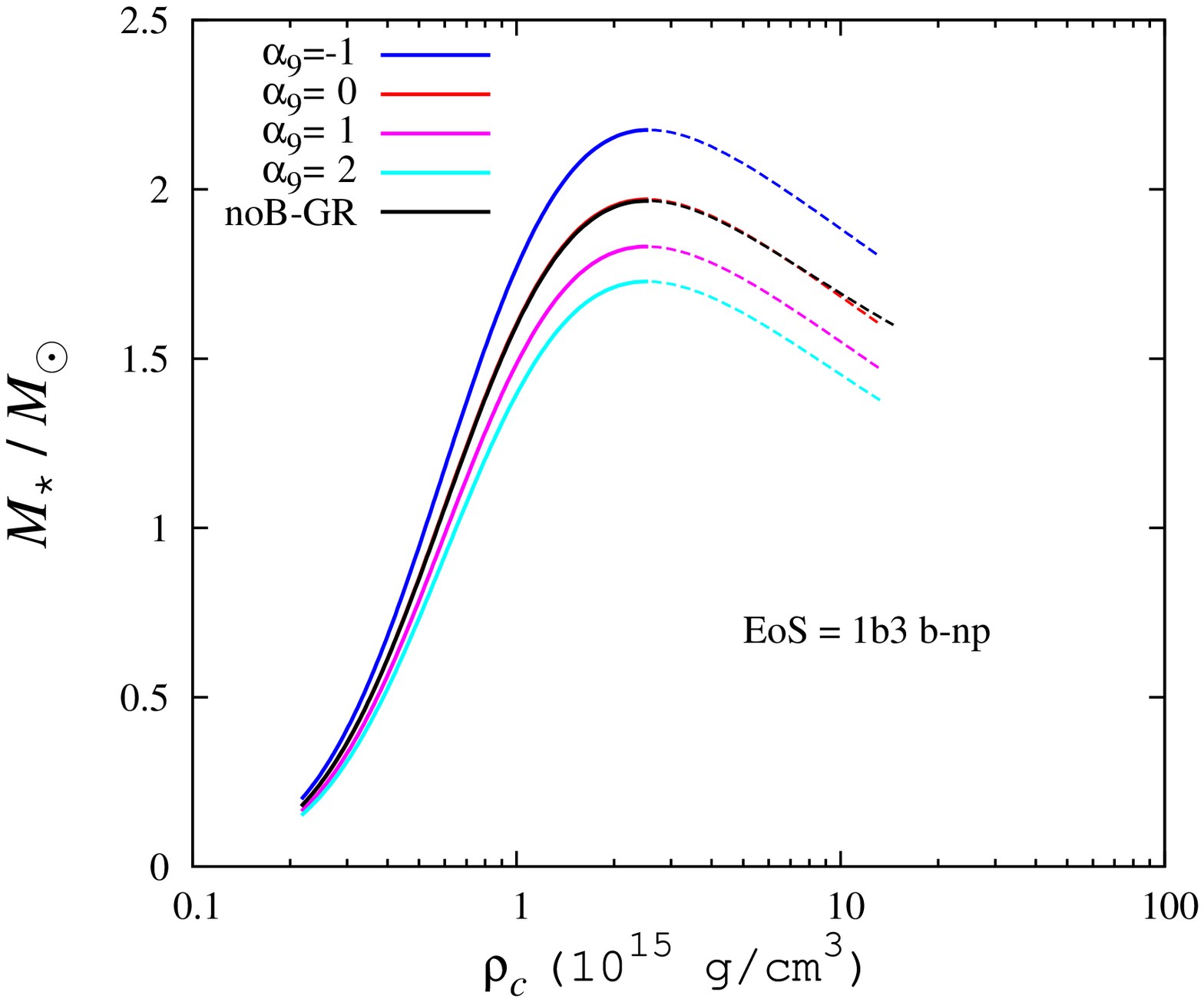}\\
\includegraphics[width=5.5cm]{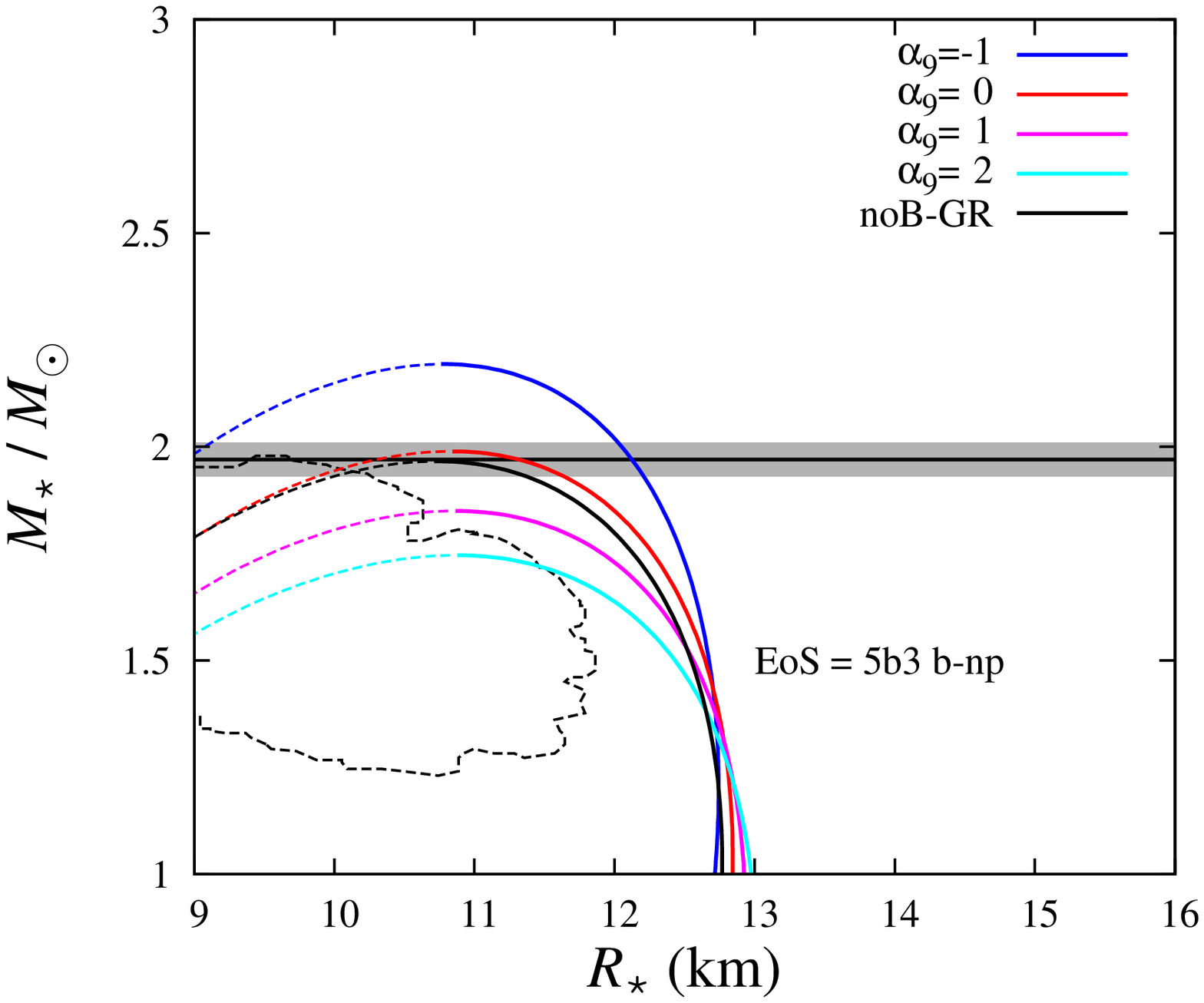}
\includegraphics[width=5.5cm]{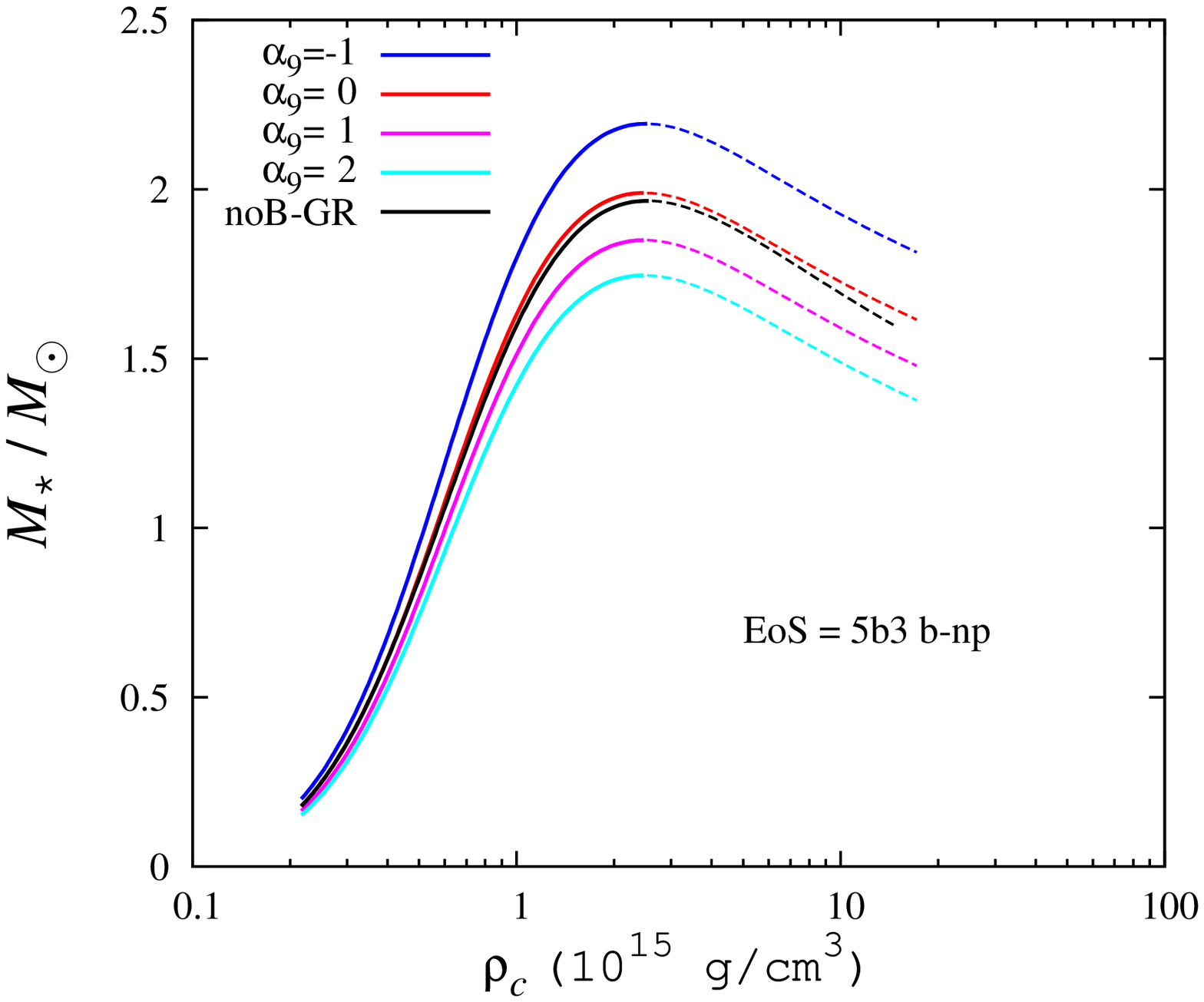}
\caption{(Color online) Mass-Radius relation (left panels) and the mass vs. central density relations (right panels) corresponding to an EoS with a $np$ phase and magnetic fields with $B_0^* = 10^{2 \sim 3}$ in the modified gravity model characterized by values for the parameter $\alpha$. Here "EoS = nbm b-np" stands for the EoS with $B_0^* = n \times 10^{m}$ in a $np$ phase. The observational constraint  \cite{Ozel10} on the  mass-radius relation is shown by the  thin dashed black contours. The mass of $M = 1.97 \pm 0.04 M_\odot$ for PSR J1614-2230 \cite{demorest2010} is shown as the horizontal black line with grey error bars.}
\label{fig1}
\end{figure}
\begin{figure}
\includegraphics[width=6.5cm]{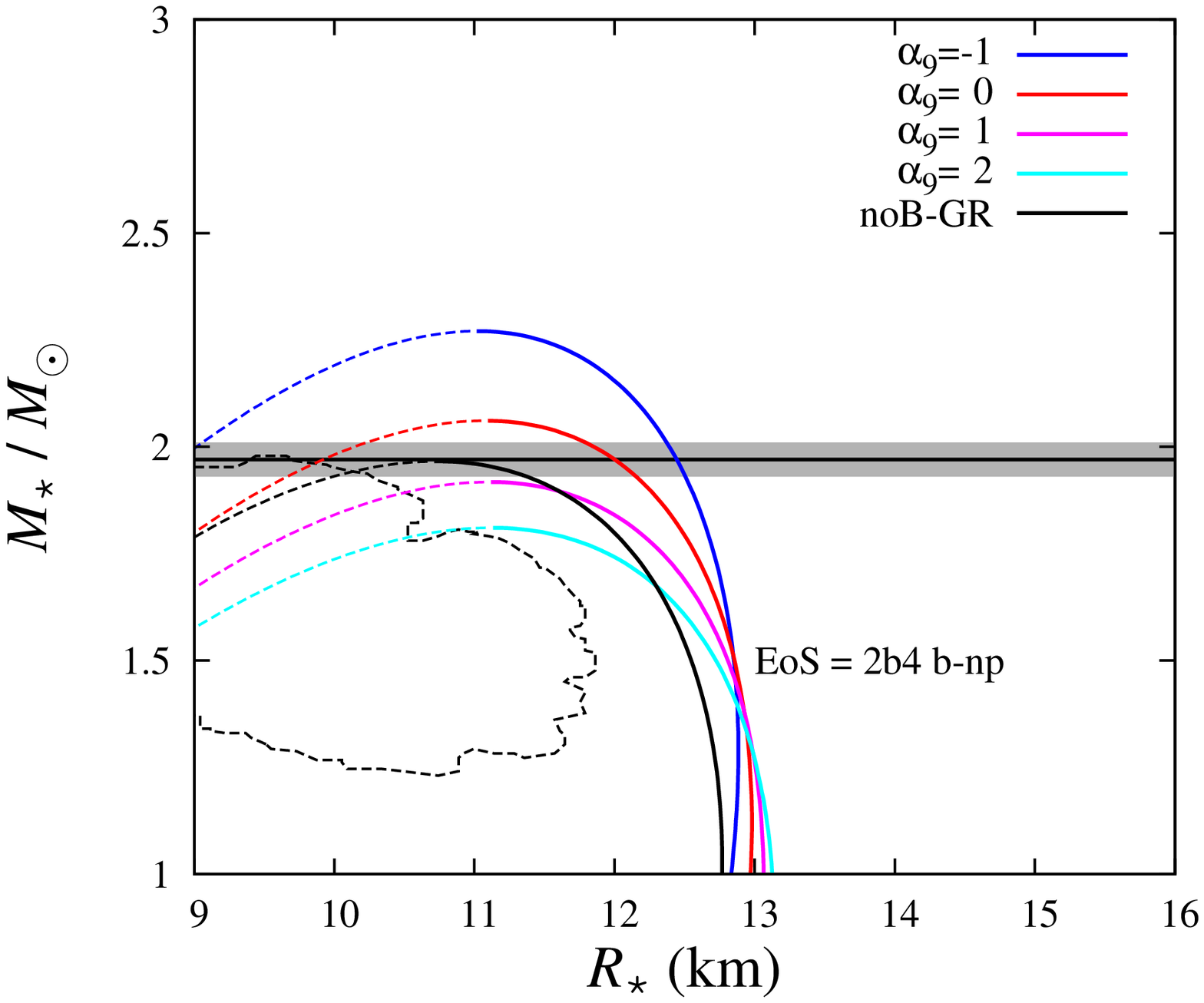}
\includegraphics[width=6.5cm]{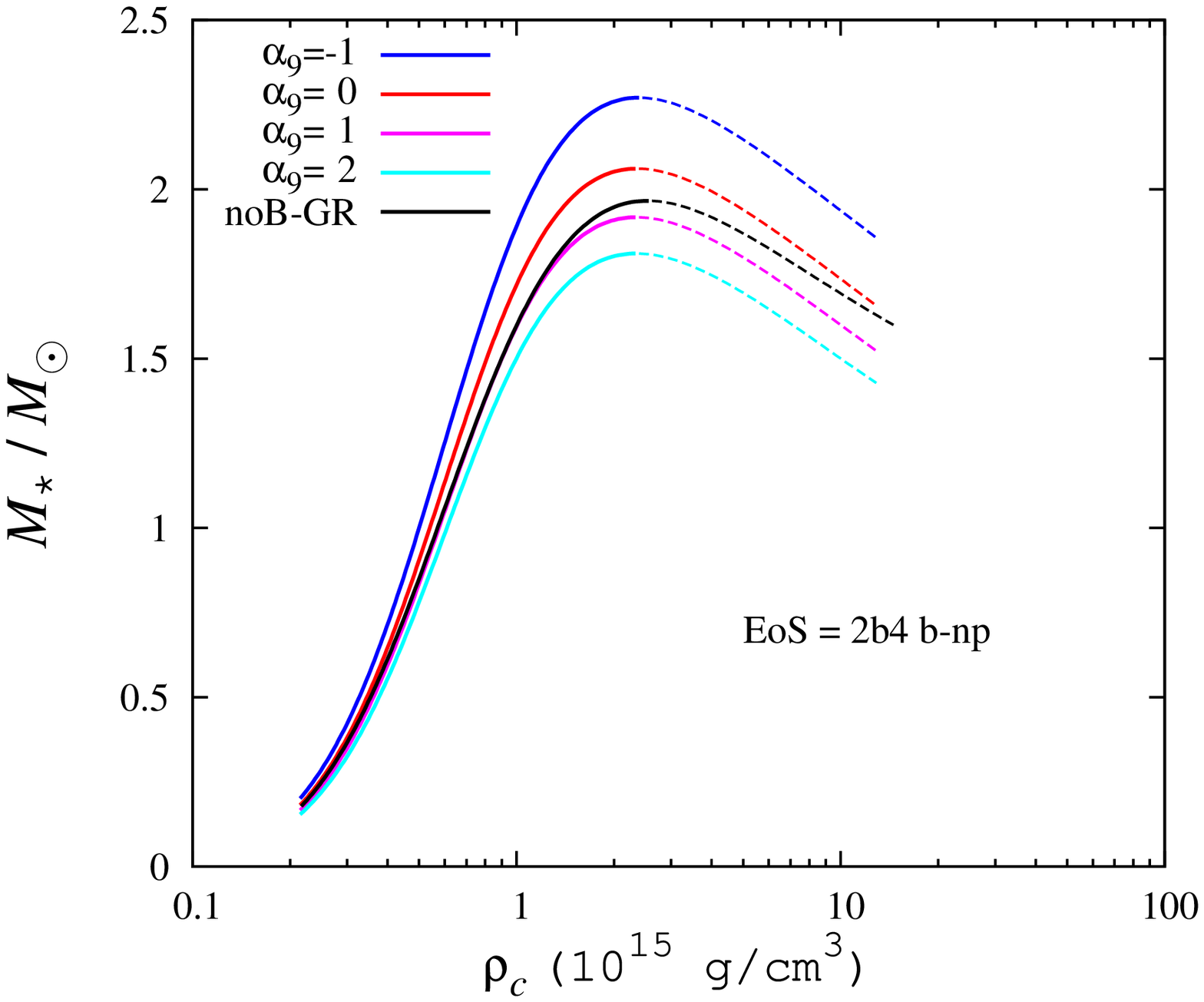}\\
\includegraphics[width=6.5cm]{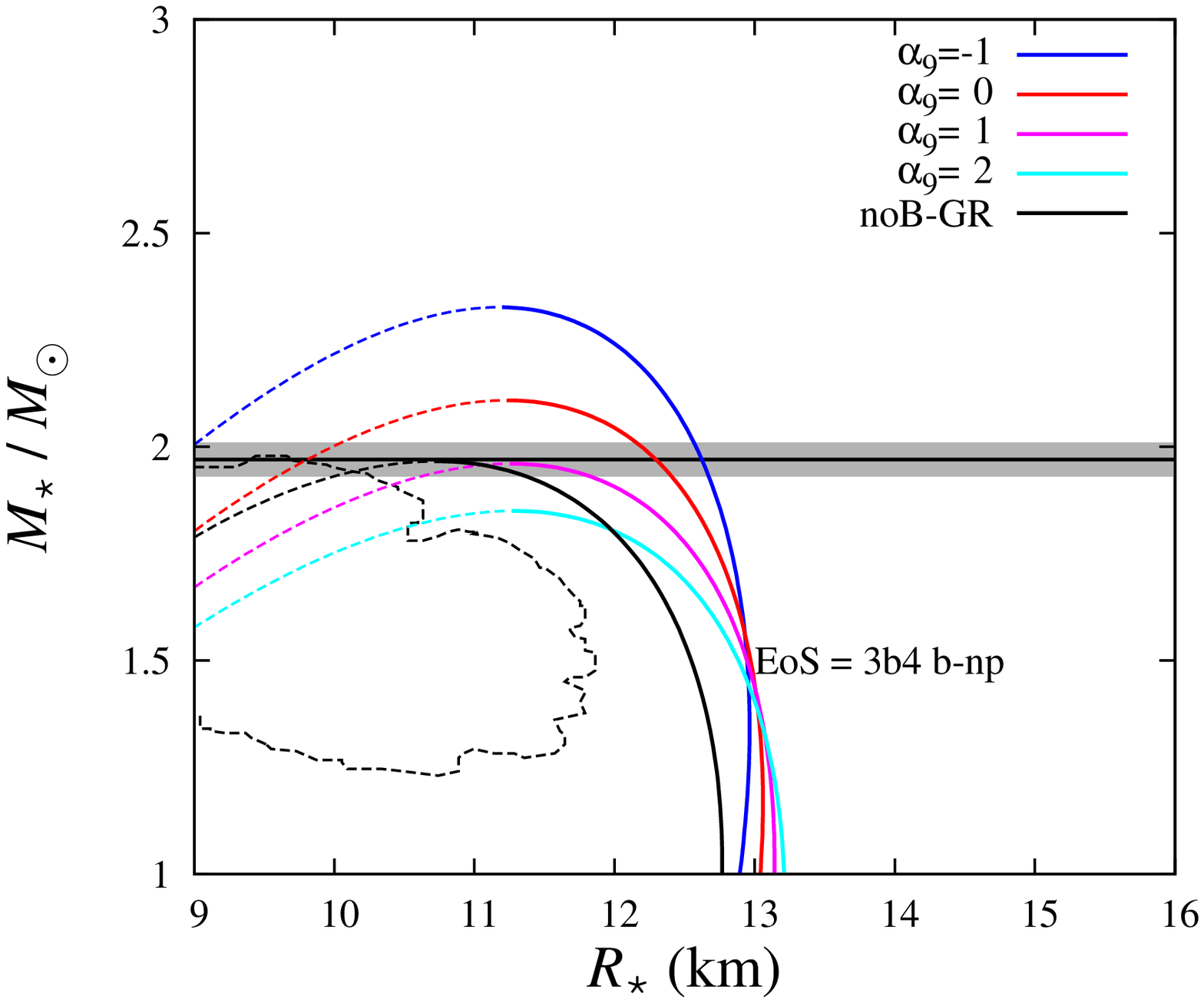}
\includegraphics[width=6.5cm]{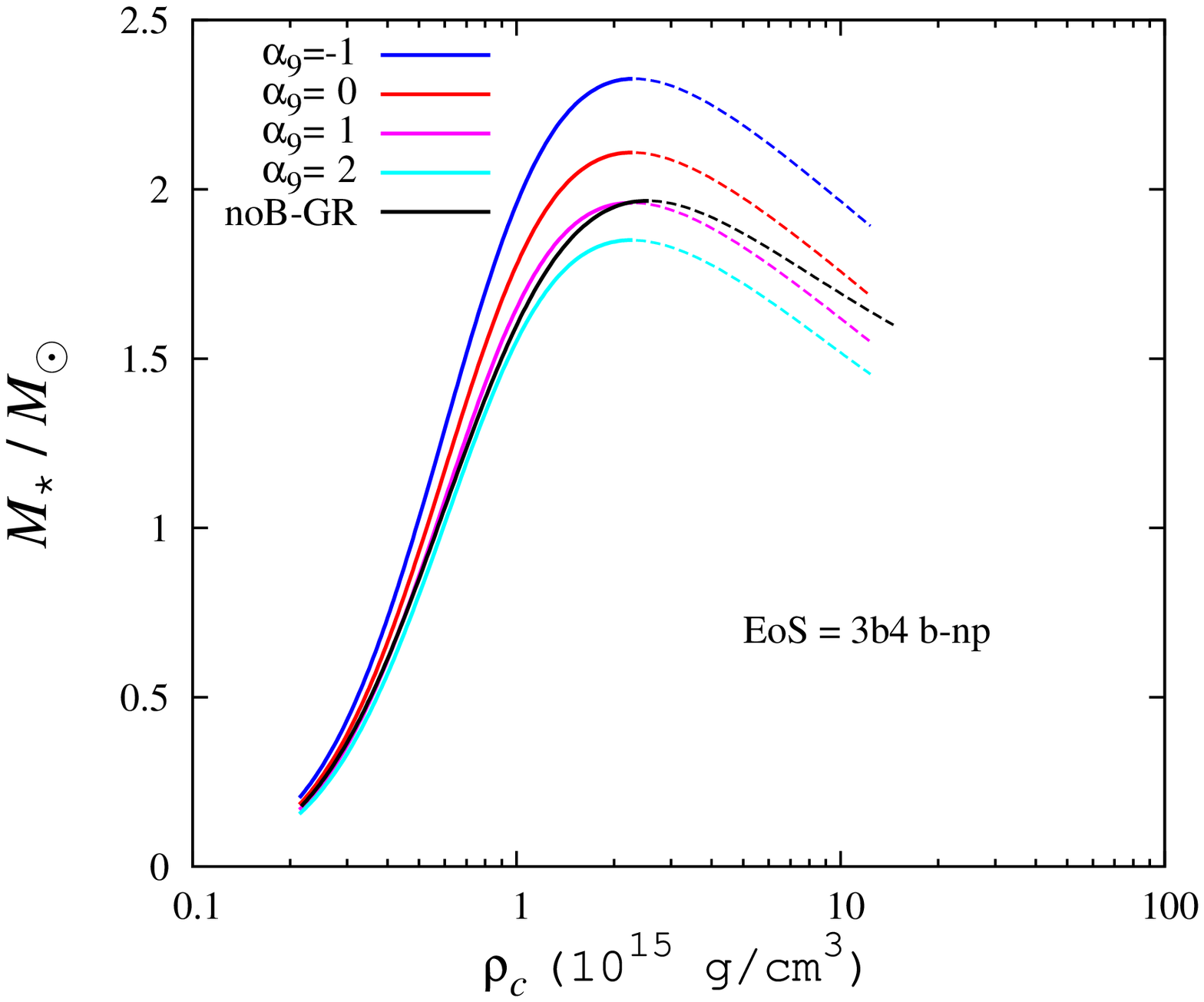}\\
\includegraphics[width=6.5cm]{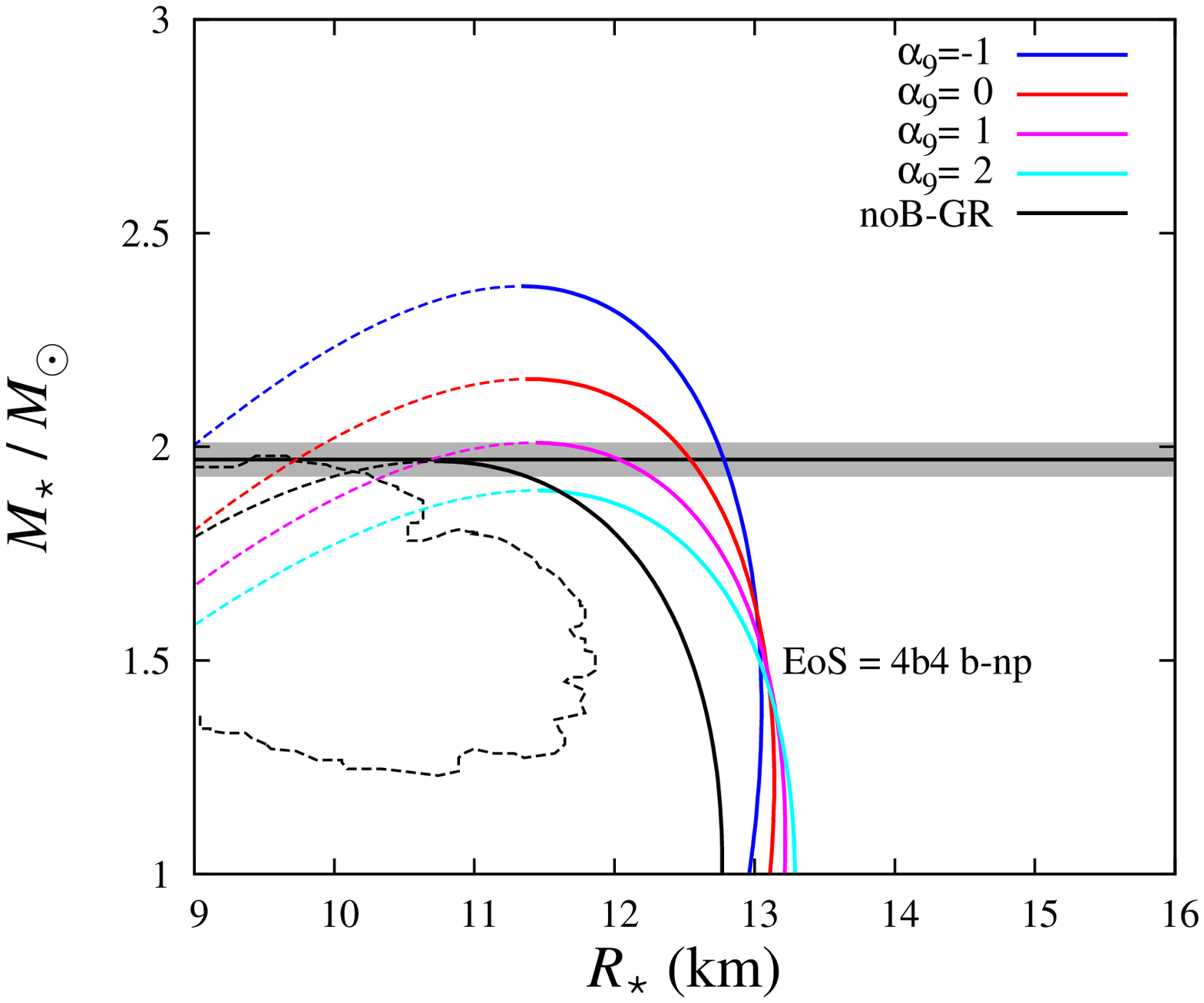}
\includegraphics[width=6.5cm]{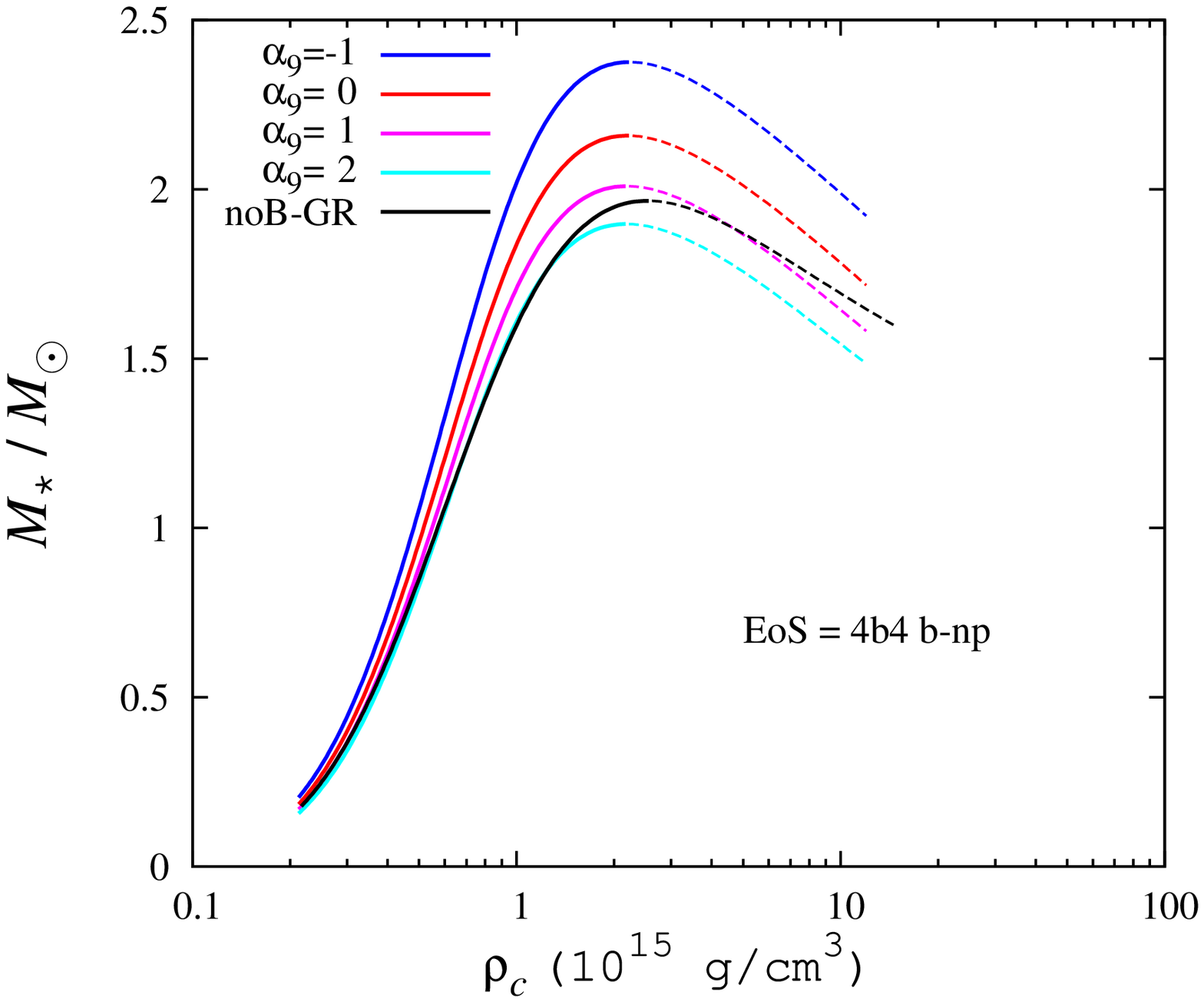}\\
\includegraphics[width=6.5cm]{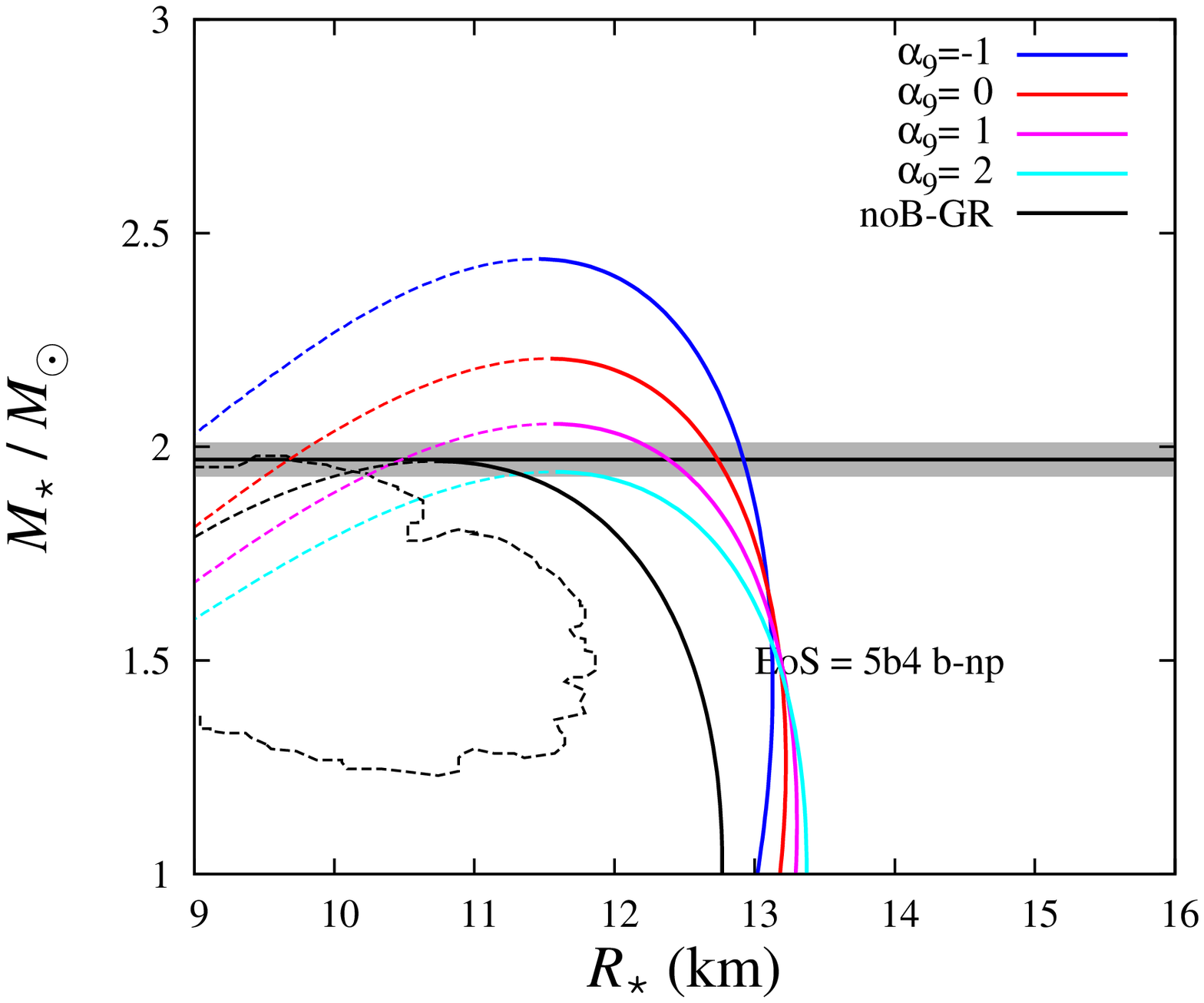}
\includegraphics[width=6.5cm]{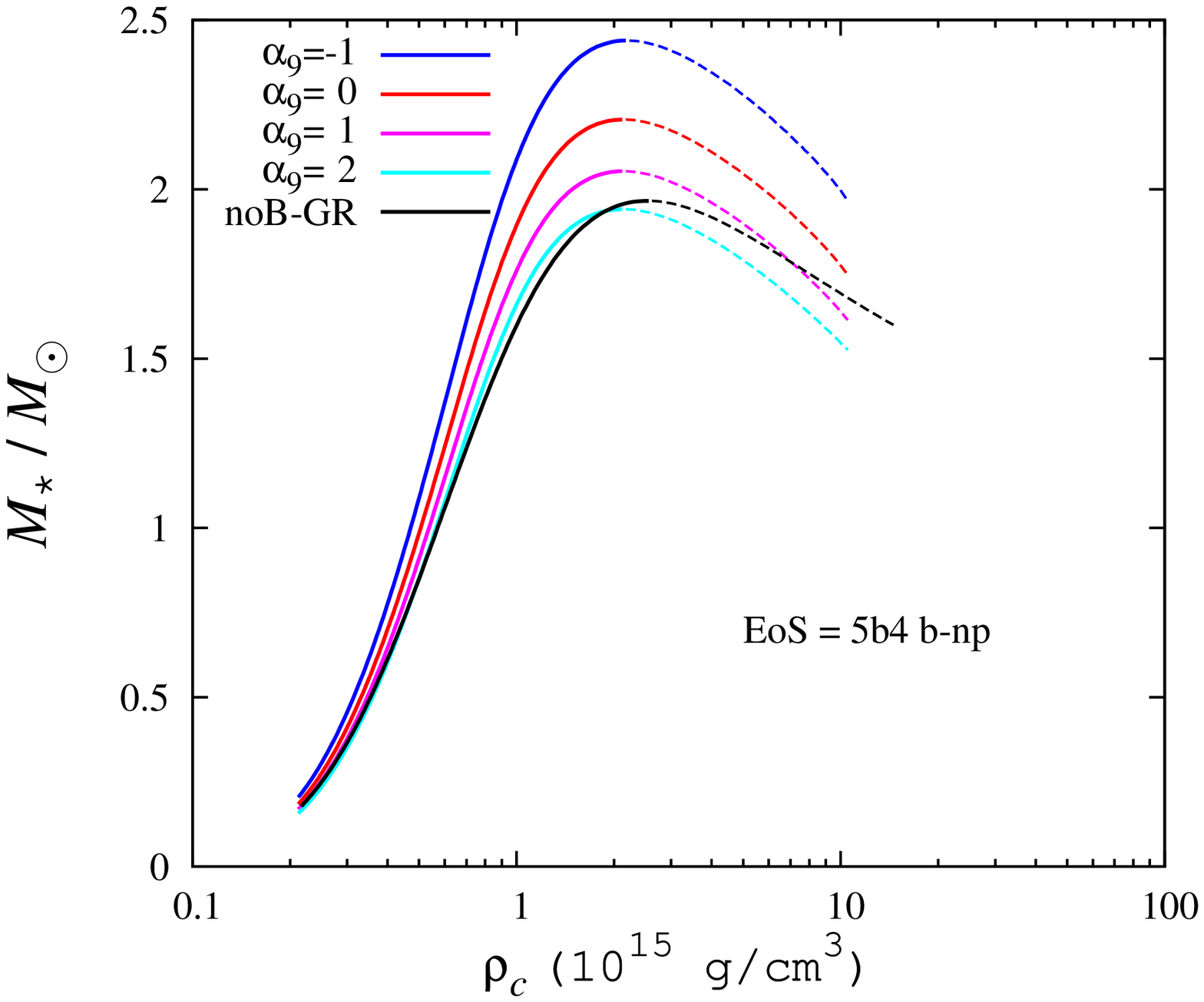}
\caption{(Color online) Same as Fig.1 but with magnetic fields $B_0^* = (2 \sim 5) \times 10^{4}$.}
\label{fig2}
\end{figure}
%
%
%
%
\begin{figure}
\includegraphics[width=6.5cm]{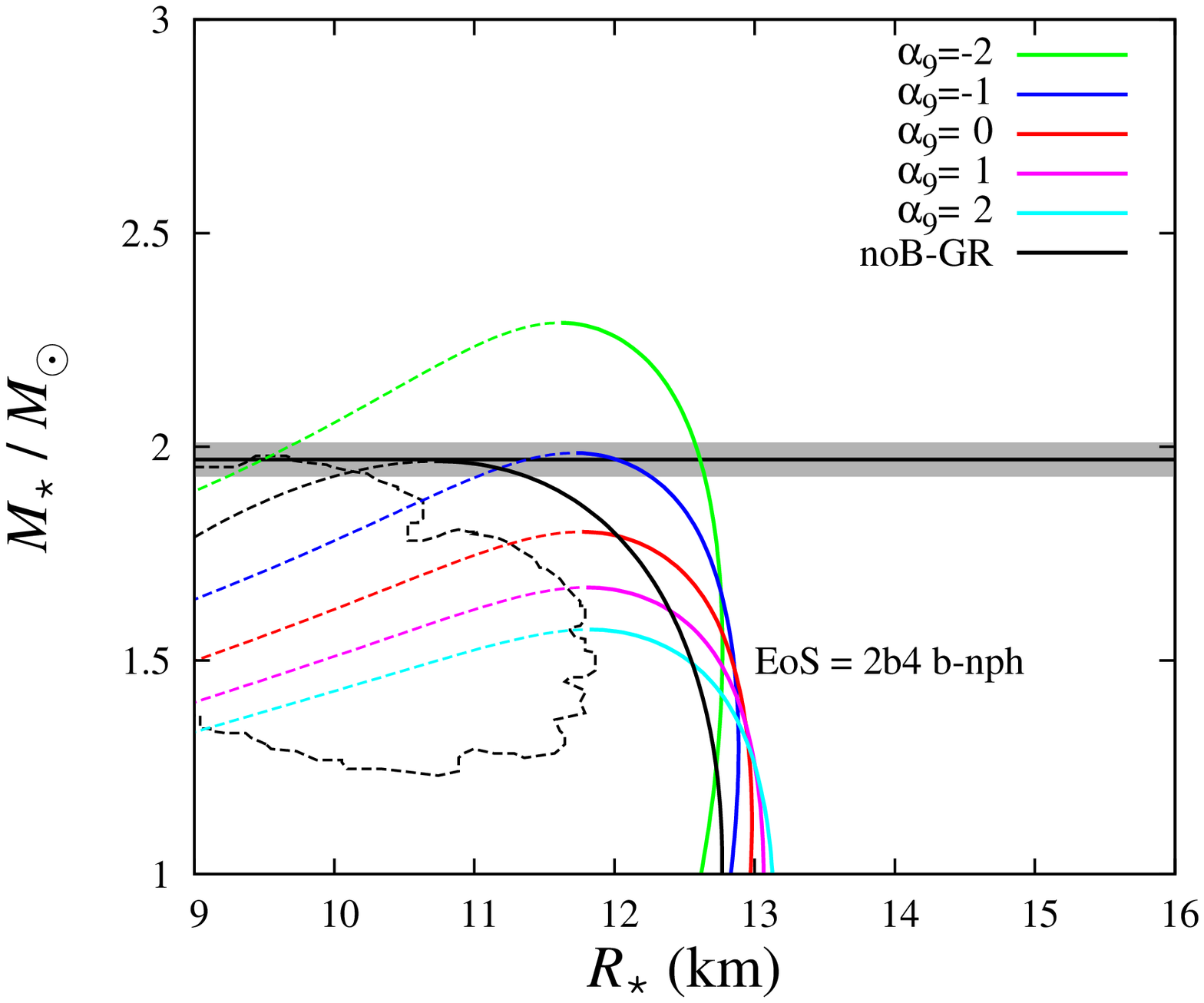}
\includegraphics[width=6.5cm]{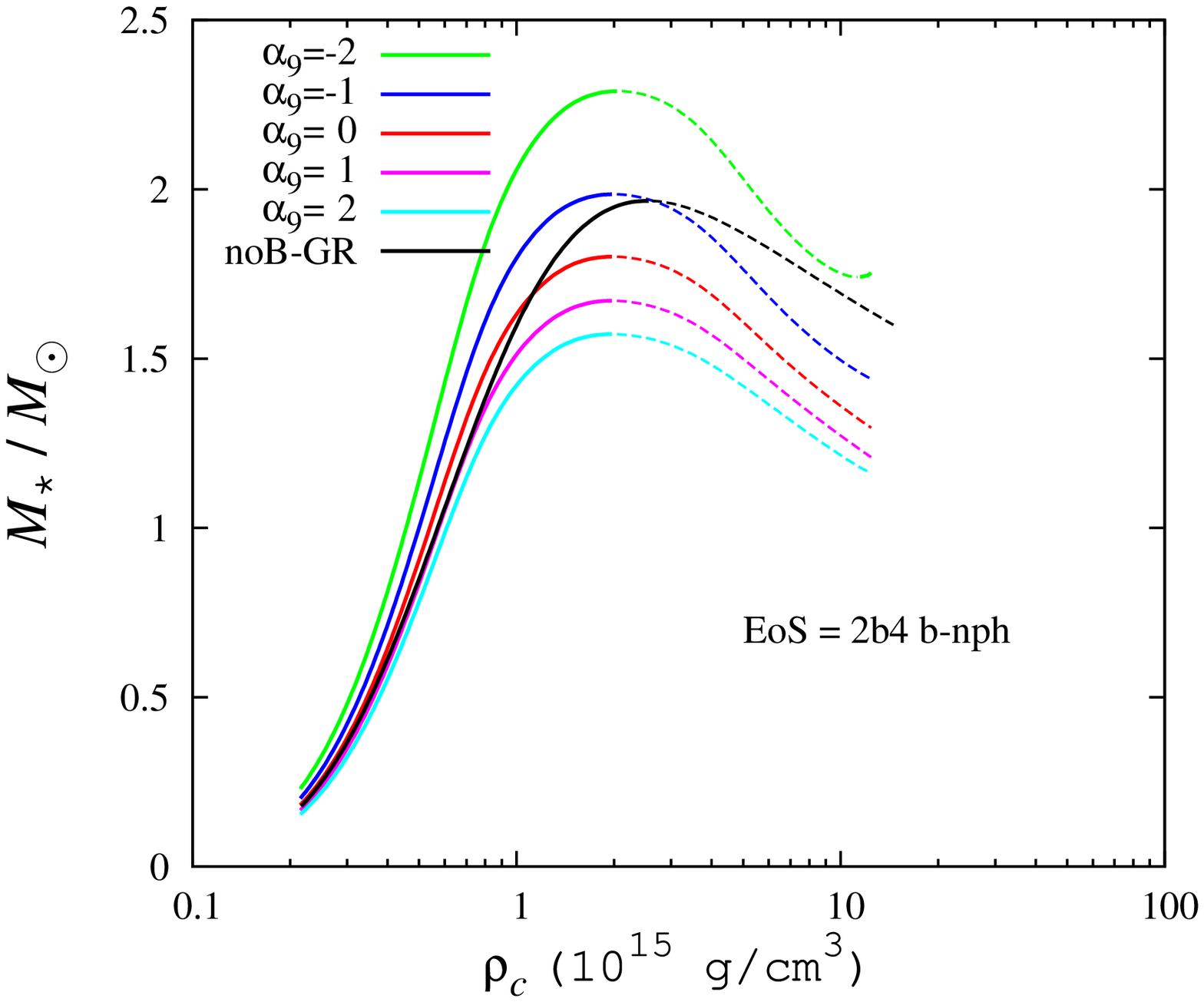}\\
\includegraphics[width=6.5cm]{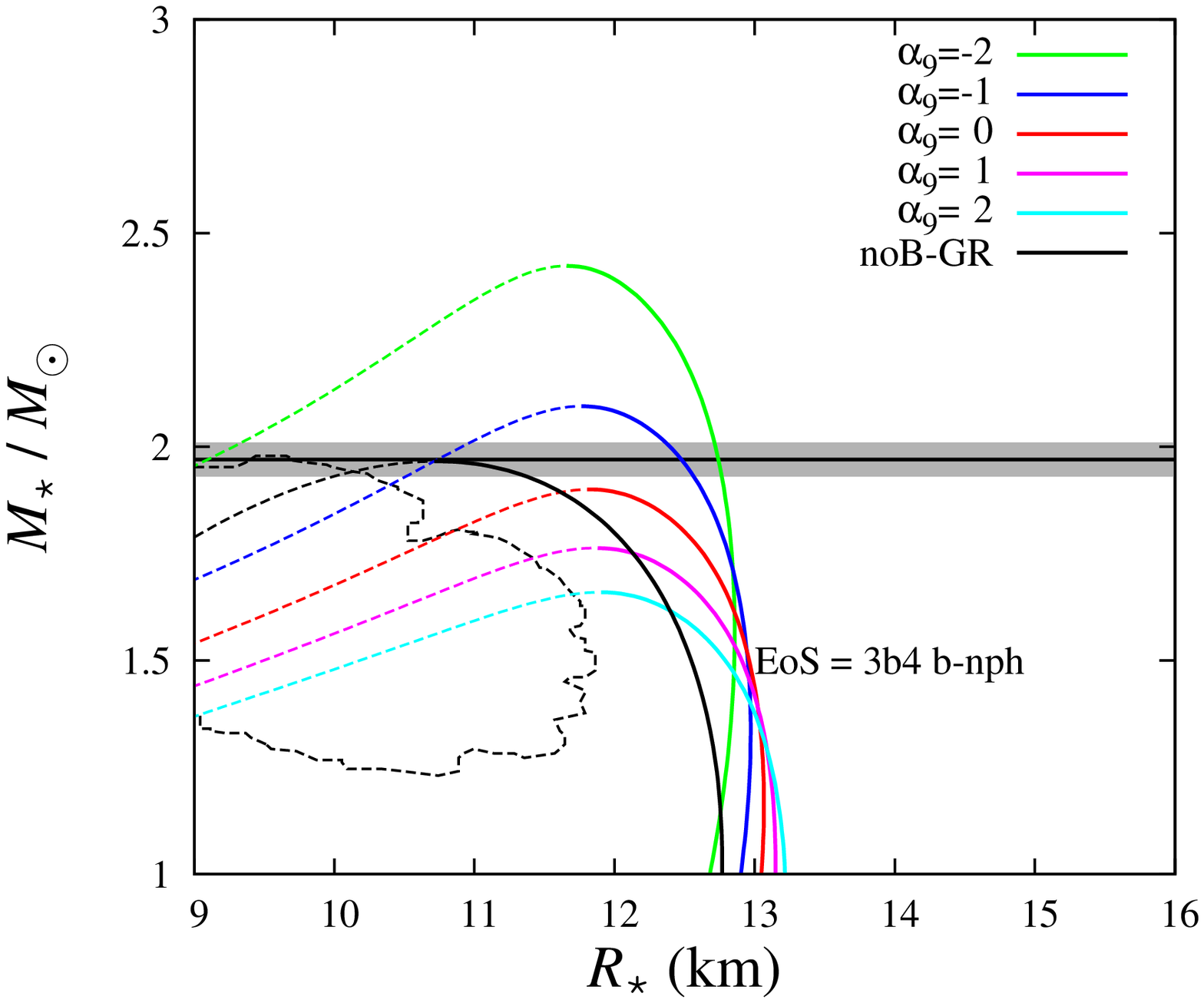}
\includegraphics[width=6.5cm]{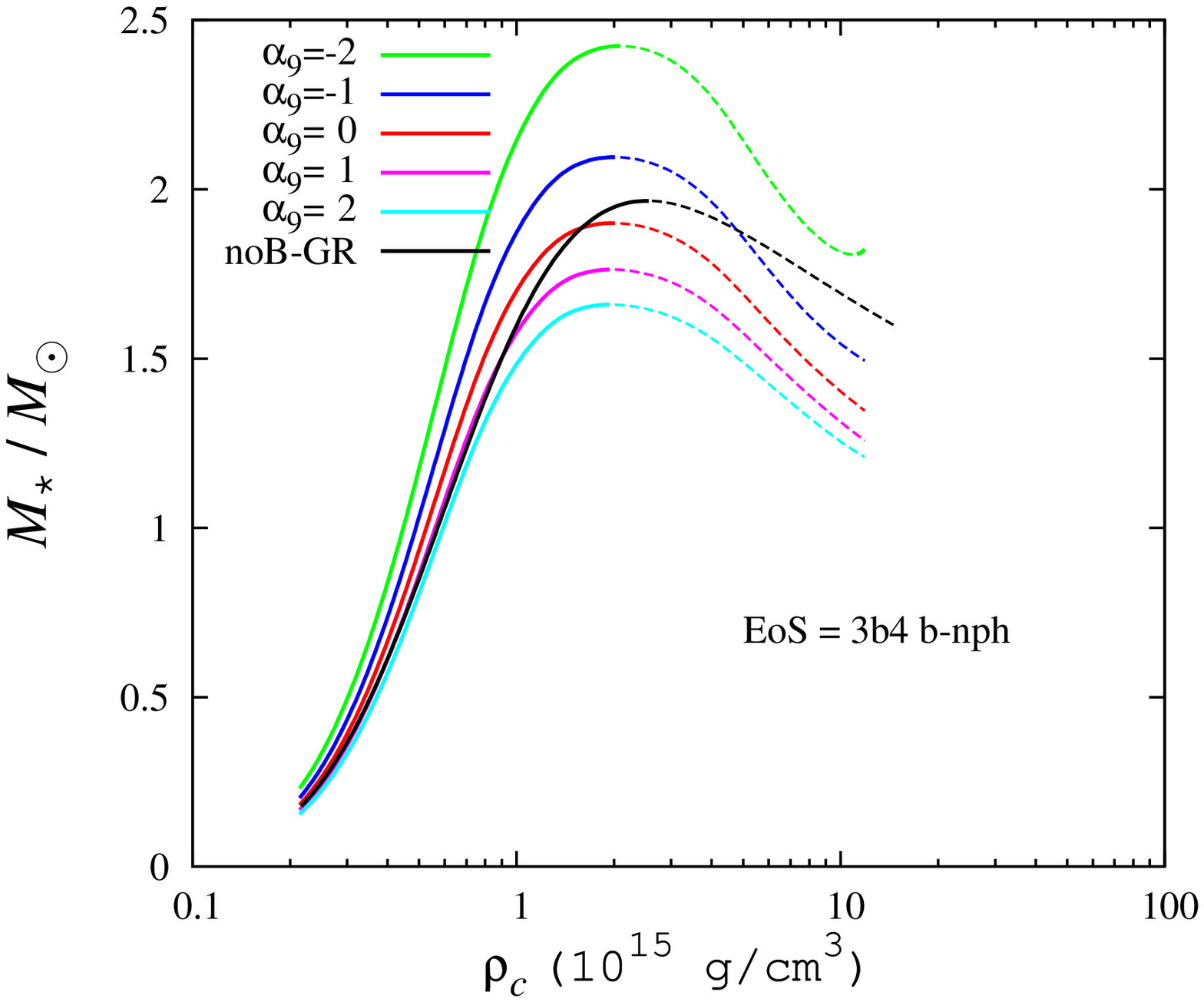}\\
\includegraphics[width=6.5cm]{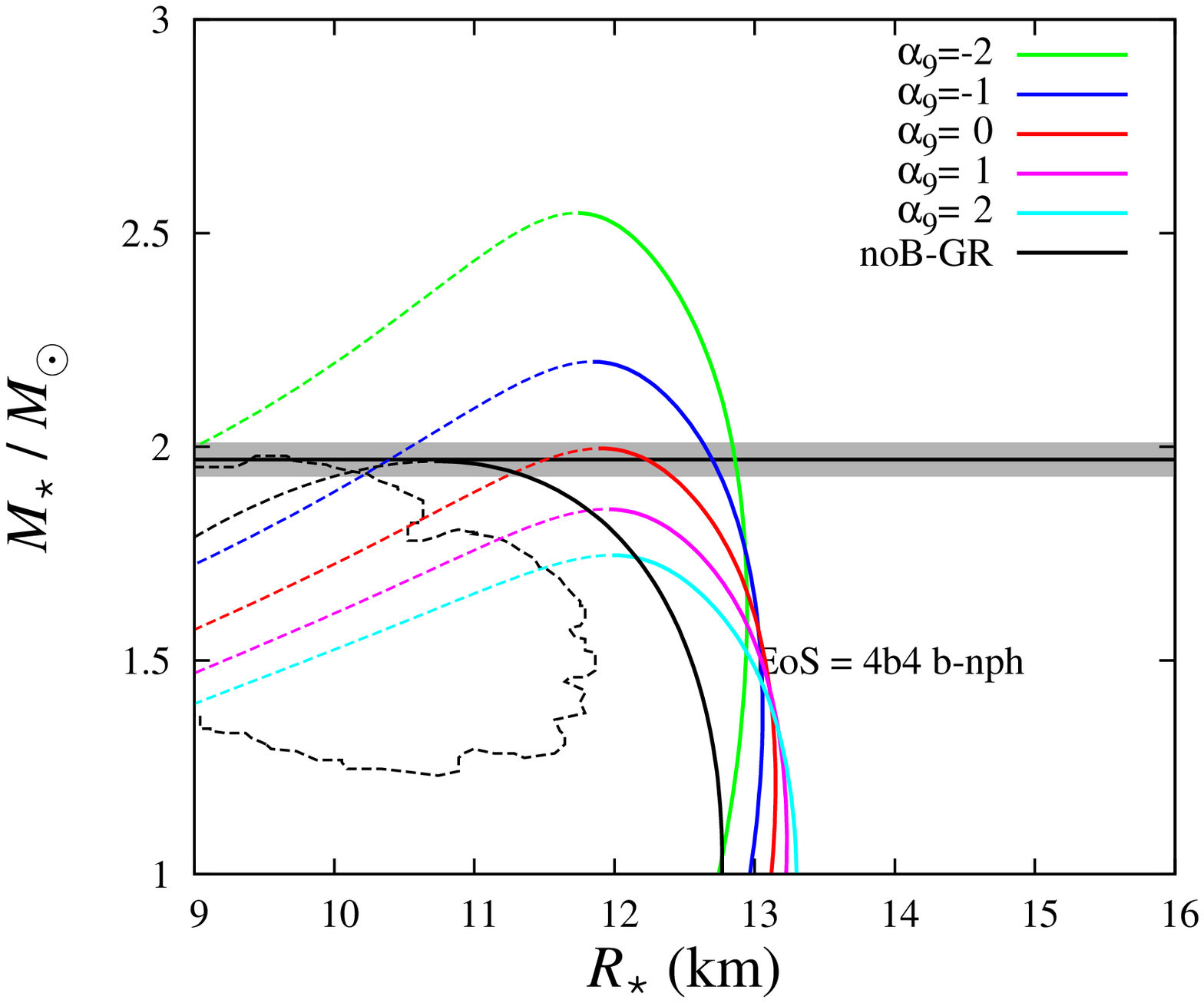}
\includegraphics[width=6.5cm]{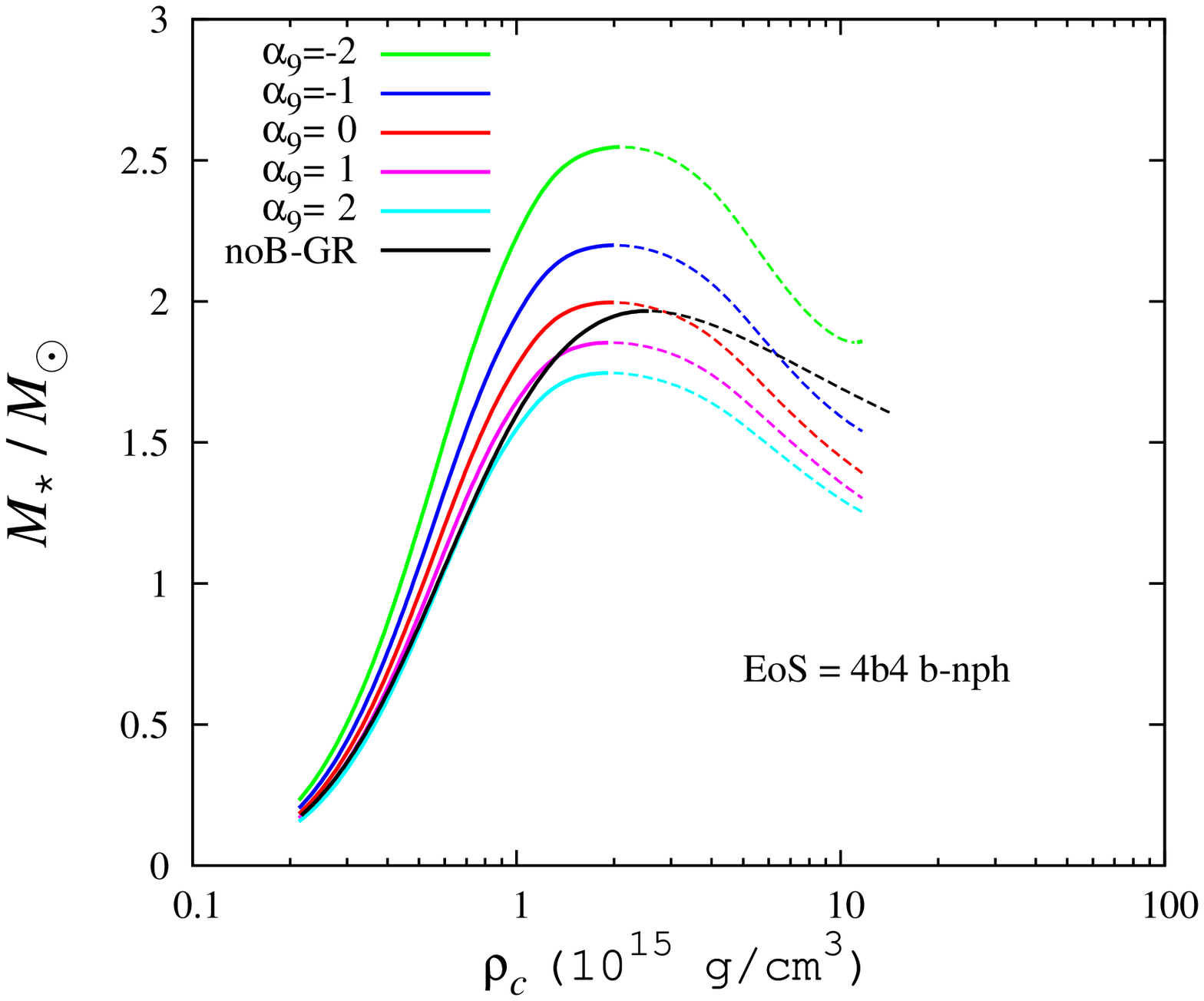}\\
\includegraphics[width=6.5cm]{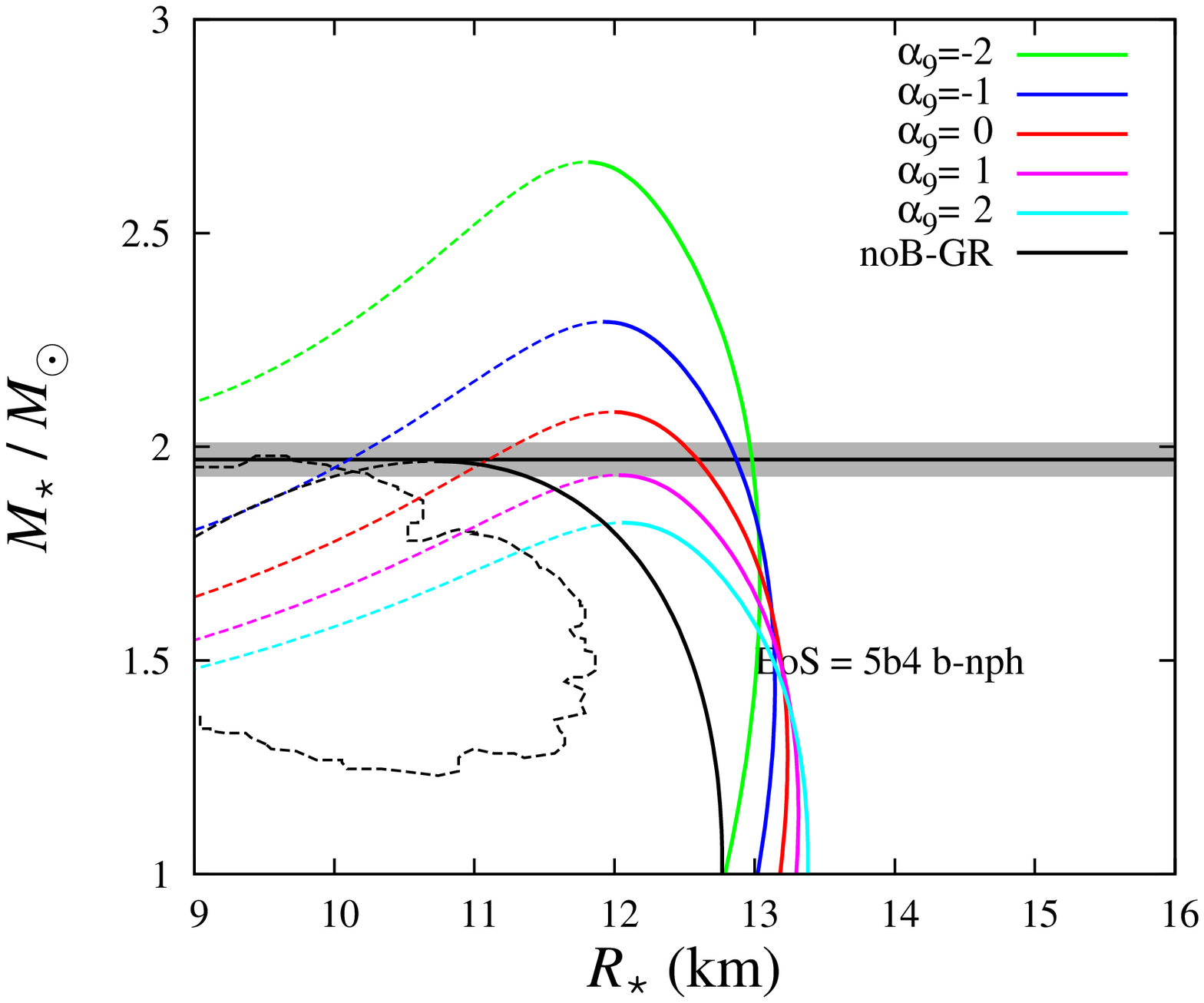}
\includegraphics[width=6.5cm]{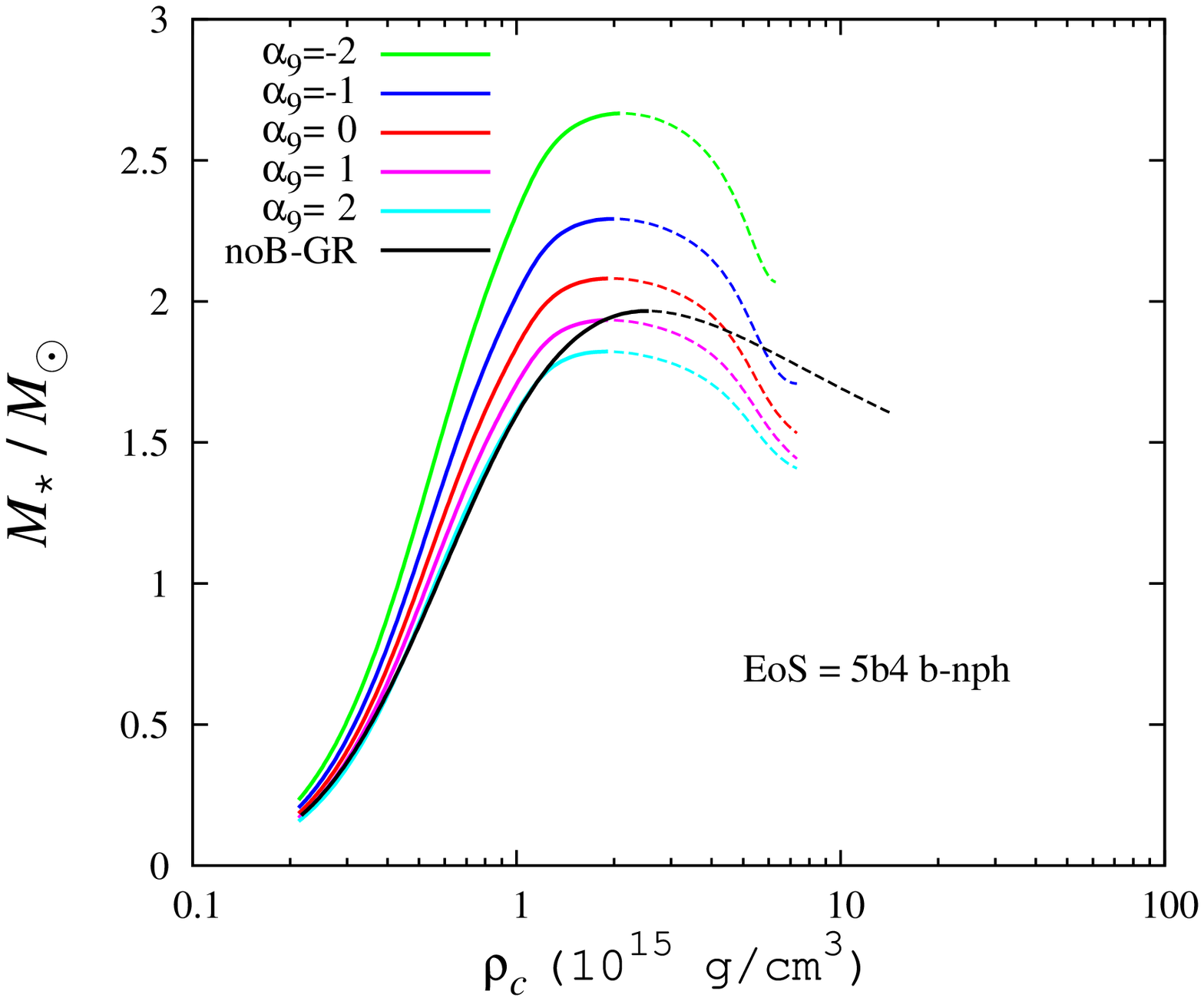}
\caption{(Color online)  Same as Fig.2 but corresponding to an EoS with a $nph$ phase, {\it i.e.} including hyperons. }
\label{fig4}
\end{figure}

\end{document}